\documentclass{aastex}
\usepackage{emulateapj5}
\usepackage{apjfonts}
\usepackage{epsf}
\slugcomment{Accepted for publication in ApJ; UTAP-417}
\shorttitle{FRACTION OF DARK HALO LENSES}
\shortauthors{OGURI}
\begin{document}
%%%%%%%%%%%%%%%%%%%%%%%%%%%%%%%%%%%%%%%%%%%%%%%%%%%%%%%%%%%%%%%%%%%%%%
%%%%%%%%%%%%%%%%%%%%%%%%%%%%%%%%%%%%%%%%%%%%%%%%%%%%%%%%%%%%%%%%%%%%%%
\title{Constraints on the Baryonic Compression and
Implications for the Fraction of Dark Halo Lenses}
%%%%%%%%%%%%%%%%%%%%%%%%%%%%%%%%%%%%%%%%%%%%%%%%%%%%%%%%%%%%%%%%%%%%%%
%%%%%%%%%%%%%%%%%%%%%%%%%%%%%%%%%%%%%%%%%%%%%%%%%%%%%%%%%%%%%%%%%%%%%%
%
%%%%%%%%%%%%%%%%%%%%%%%%%%%%%%%%%%%%%%%%%%%%%%%%%%%%%%%%%%%%%%%%%%%%%%
\author{Masamune Oguri}
\affil{Department of Physics, School of Science, University of Tokyo,
Tokyo 113-0033, Japan.} 
\email{oguri@utap.phys.s.u-tokyo.ac.jp}
%%%%%%%%%%%%%%%%%%%%%%%%%%%%%%%%%%%%%%%%%%%%%%%%%%%%%%%%%%%%%%%%%%%%%%
%
\received{2002 May 14}
\accepted{2002 July 23}
\begin{abstract}
We predict the fraction of dark halo lenses, that is, the fraction of
lens systems produced by the gravitational potential of dark halos, on
the basis of a simple parametric model of baryonic compression. The 
fraction of dark halo lenses primarily contains information on the
effect of baryonic compression and the density profile of dark
halos, and is expected to be insensitive to cosmological parameters 
and source population. The model we adopt comprises the galaxy formation
probability $p_{\rm g}(M)$ which describes the global efficiency of
baryonic compression and the ratio of circular velocities of galaxies
to virial velocities of dark halos $\gamma_v=v_{\rm c}/v_{\rm vir}$
which means how the inner structure of dark halos is modified due to
baryonic compression. The model parameters are constrained from the
velocity function of galaxies and the distribution of image separations
in gravitational lensing, although the degeneracy between model
parameters still remains. We show that the fraction of dark halo lenses
depends strongly on $\gamma_v$ and the density profile of dark halos
 such as inner slope $\alpha$. This means that the observation of
 the fraction of dark halos can break the degeneracy between model
 parameters if the density profile of dark halo lenses is fully settled. On
 the other hand, by restricting $\gamma_v$ to physically plausible range
 we can predict the lower limit of the fraction of dark halo lenses on
 the basis of our model. Our result indicates that steeper inner cusps
 of dark halos  ($\alpha\gtrsim 1.5$) or too centrally concentrated dark
 halos are inconsistent with the lack of dark halo lenses in
 observations.
\end{abstract} 
%
%%%%%%%%%%%%%%%%%%%%%%%%%%%%%%%%%%%%%%%%%%%%%%%%%%%%%%%%%%%
%%%%%%%%%%%%%%%%%%%%%%%%%%%%%%%%%%%%%%%%%%%%%%%%%%%%%%%%%%%
\section{Introduction}
%%%%%%%%%%%%%%%%%%%%%%%%%%%%%%%%%%%%%%%%%%%%%%%%%%%%%%%%%%%
%%%%%%%%%%%%%%%%%%%%%%%%%%%%%%%%%%%%%%%%%%%%%%%%%%%%%%%%%%%
Strong gravitational lensing offers a powerful probe of the matter
distribution in the universe. So far $\sim60$ lensed quasars are known,
and their properties are summarized by CfA/Arizona Space Telescope Lens
Survey (CASTLES\footnote{Kochanek, C.~S., Falco, E.~E., Impey, C.,
Lehar, J., McLeod, B., \& Rix, H.-W., http://cfa-www.harvard.edu/castles/}).
Most of these lens systems have image separations $\theta\lesssim3''$.
The cold dark matter (CDM) scenario, however, predicts 
sufficiently cuspy dark halos \citep*[e.g.,][]{navarro96,navarro97},  
thus dark halos can produce the significant amount of multiple images
even at $\theta\gtrsim 3''$
\citep*{wyithe01,keeton01b,takahashi01,li02,oguri02a}. It has been unclear
whether the distribution of image separations should be computed based
on galaxies \citep*[using the luminosity function and the density profile of
galaxies;][]{turner84,turner90,fukugita91,fukugita92,maoz93,kochanek96,chiba99}
or dark halos \citep*[using the mass function and the density profile  
of dark
halos;][]{narayan88,cen94,wambsganss95,kochanek95,maoz97,wambsganss98,mortlock00,wyithe01,keeton01b,takahashi01,oguri02a}.
The distribution of image separations, however, clearly indicates that
the use of only dark matter properties cannot match observations. That
is, the observed distribution of image separations is never reproduced
from the theoretical calculation using the mass function of dark halos
and only one population for lensing objects \citep[e.g.,][]{li02}. 
The possible solution to explain  the distribution of image separations
is the modification of inner structure of dark halos by introducing
baryonic cooling
\citep*{keeton98,porciani00,kochanek01b,keeton01a,sarbu01,li02}. In this
picture, inside low mass halos baryons are efficiently compressed and
form sufficiently concentrated galaxies, while inside larger halos such
as group- or cluster-mass halos global baryon cooling hardly occurs and
thus the inner structure of dark halos remains unmodified.  

Then a question comes to our mind: {\it what is the fraction of dark
halo lenses?} 
Here the term ``dark halo lenses'' is used to describe lenses which are
produced by the gravitational  potential of dark halos. In other words,
the lens objects of dark halo lenses are dark halos 
which have no central galaxies or have central galaxies but they are too
small to dominate in gravitational lensing. Dark halo lenses  exhibit
characteristic properties such as the small flux ratios and the
detectable odd images \citep{rusin02}, thus can be distinguished from
usual galaxy lenses. 
Since dark halos are expected to have a steep central cusp, the
significant amount of dark halo lenses should be observed. But so far no
confirmed dark halo lens system is observed in strong gravitational
lensing survey. The exception is arc statistics in rich clusters
\citep*{bartelmann98,williams99,meneghetti01,molikawa01,oguri01,oguri02b}, 
but the known cluster lenses were all found by searching for lenses in
detail after identifying a rich cluster. In the surveys which first
identify source objects and see whether they are lensed or not, it seems
that dark halo lenses have not been observed yet: statistical argument
\citep*{kochanek99} 
and individual properties \citep{rusin02} imply that current ambiguous 
quasar pairs are likely to be binary quasars. Even known lensing systems
in clusters, such as Q0957+561 \citep*{walsh79}, are produced mainly by
a galaxy in the cluster. The cluster potential contributes to lensing
only as a perturbation. Therefore, it should be checked whether the lack
of dark halo lenses in observations really reconciles with the
theoretical prediction. 

Description of baryonic effects needs detailed models for the star
formation and feedback \citep[e.g.,][]{cole00}. Instead, in this paper
we predict the expected fraction of dark halo lenses on the basis  
of a simple (minimal) parametric model \citep{kochanek01a}. This model 
comprises the formation probability of galaxies, $p_{\rm g}(M)$, and 
the ratio of circular velocities of galaxies to virial velocities of dark 
halos, $\gamma_v$. The former describes the global efficiency of baryonic 
compression, and the latter models the modification of inner structure of dark 
halos due to baryonic compression. The model parameters are chosen so 
as to reproduce the velocity function of galaxies and the distribution of 
image separations. Although there remains the strong degeneracy between model 
parameters, this degeneracy can be broken from the observation of the
fraction of dark halos if we fix the density profile of dark halos. On
the other hand, by restricting a range of $\gamma_v$ from various theories and 
observations, we can also derive the lower limit of the fraction of dark
halo lenses. Our main finding is that steep inner slopes of dark halos 
($\alpha\gtrsim1.5$) or too centrally concentrated dark halos are
inconsistent with the lack of dark halo lenses in observations, even if 
various uncertainties are taken into account. Although this constraint on the 
density profile is not so severe, a large lens sample obtained by e.g., 
Sloan Digital Sky Survey (SDSS) can put tighter constraints on the 
density profile of dark halos as well as the model of baryonic
compression. 

The plan of this paper is as follows. In \S \ref{sec:th}, we describe the 
model of baryonic compression. Section \ref{sec:const} is devoted to 
constrain the model parameters, and \S \ref{sec:frac} presents our 
predictions for the fraction of dark halo lenses. Finally, we summarize
conclusions in \S \ref{sec:conc}. Throughout this paper, we assume 
the lambda-dominated cosmology 
$(\Omega_0, \lambda_0,h,\sigma_8)=(0.3,0.7,0.7,1.04)$, where 
the Hubble constant in units of $100{\rm km\,s^{-1}Mpc^{-1}}$ is denoted by
$h$. As shown below, however, our results are quite insensitive to a 
particular choice of cosmological parameters.

%%%%%%%%%%%%%%%%%%%%%%%%%%%%%%%%%%%%%%%%%%%%%%%%%%%%%%%%%%%
%%%%%%%%%%%%%%%%%%%%%%%%%%%%%%%%%%%%%%%%%%%%%%%%%%%%%%%%%%%
\section{Theoretical Model\label{sec:th}}
%%%%%%%%%%%%%%%%%%%%%%%%%%%%%%%%%%%%%%%%%%%%%%%%%%%%%%%%%%%
%%%%%%%%%%%%%%%%%%%%%%%%%%%%%%%%%%%%%%%%%%%%%%%%%%%%%%%%%%%

\subsection{Effects of Baryons on Dark Halos\label{sec:th:baryon}}

The obvious difference between velocity functions of galaxies and 
dark halos at high velocity \citep[e.g.,][]{gonzalez00} indicates that 
the efficiency of the baryon compression changes from galaxy-mass scale
to group- and cluster-mass scales. If the mass of dark halos is
sufficiently large, the baryon cooling time $\tau_{\rm cool}$ which
increases with halo mass \citep[e.g.,][]{cole00} becomes larger than the
age of dark halos which slightly deceases with halo mass
\citep[e.g.,][]{lacey93,kitayama96}. Thus baryons inside such a massive
dark halo do not form so large galaxies as to collect most of baryons in
the dark halo. This yields a steep cutoff at high velocity in the
galaxy velocity function. To model this, following \citet{kochanek01a},
we introduce the probability  $p_{\rm g}(M)$ that a sufficiently large
galaxy which collects most of internal baryons is formed inside 
a halo of mass $M$. For example, galaxy-mass halos with mass $M$ usually
have large galaxies with mass $M_{\rm gal}\sim(\Omega_{\rm
b}/\Omega_0)M$. On the other hand, groups or clusters also have galaxies
but their mass are small even for galaxies which lie at the center of
halos, $M_{\rm gal}\ll(\Omega_{\rm b}/\Omega_0)M$, here in this case 
$M\gtrsim10^{13}h^{-1}M_\odot$ \citep[e.g.,][]{yoshikawa01}. Therefore
we regard galaxies in groups or clusters as ``substructures'' and do not
take into account. We adopt a parametric model \citep{kochanek01a}
\begin{eqnarray}
 p_{\rm g}(M)&=&
\left\{
\begin{array}{@{\hspace{0.6mm}}ll}
\displaystyle{1} & \mbox{($M<M_{\rm h}$)},\\
\displaystyle{\exp\left[1-\left(\frac{M}{M_{\rm h}}\right)^{\delta_{\rm h}}\right] } & \mbox{($M>M_{\rm h}$)}.
\end{array}
\right.\label{gfp}
\end{eqnarray}
Although the overall factor of $p_{\rm g}(M)$ should become another
parameter, we neglect it. The effect of the overall factor of $p_{\rm
g}(M)$ on the fraction of dark halo lenses will be discussed in \S
\ref{sec:frac}. 

Also, there is a difference in the slopes of velocity
functions of dark halos and galaxies at small velocity 
\citep[e.g.,][]{gonzalez00,nagamine01}, and this 
demands the modification of $p_{\rm g}(M)$ at low $M$. 
We, however, neglect this effect because we are not interested in low 
velocity galaxies which hardly contribute to lensing statistics.

The baryon compression also changes inner structure of dark halos 
\citep*[e.g.,][]{blumenthal86,mo98}. Before the baryon cooling occurs, 
we assume that the density profile of dark halos is well described by 
the one-parameter family of the form \citep{zhao96,jing00a}
\begin{equation}
 \rho(r)=\frac{\rho_{\rm crit}\delta_{\rm c}}
{\left(r/r_{\rm s}\right)^\alpha\left(1+r/r_{\rm s}\right)^{3-\alpha}},
\label{nfw}
\end{equation}
where $r_{\rm s}=r_{\rm vir}/c_{\rm vir}$ and $c_{\rm vir}$ is the 
concentration parameter. We adopt the mass and redshift dependence
reported by \citet{bullock01}:
\begin{equation}
 c_{\rm vir}(M, z)=\frac{8}{1+z}\left(\frac{M}{10^{14}h^{-1}M_{\odot}}\right)^{-0.13},\label{bul}
\end{equation}
for $\alpha=1$, and generalize it to $\alpha\neq1$ by multiplying the
factor $(2-\alpha)$ \citep{keeton01b}. We also take account of scatter of
the concentration parameter which has a log-normal distribution with the
dispersion of $\sigma_{\rm c}=0.18$ \citep{jing00b,bullock01}.
The characteristic density $\delta_{\rm c}$ can be computed using the
spherical collapse model \citep[see][]{oguri01}. While the correct value
of $\alpha$ is still unclear, the existence of a cups with
$1\lesssim\alpha\lesssim1.5$ has been established in recent N-body
simulations \citep{navarro96,navarro97,moore99,ghigna00,jing00a,klypin01,fukushige01,fukushige02,power02}.

Next consider the modification of inner structure of dark halos induced
by baryonic compression. In general, cooled baryons (galaxies) are
more centrally concentrated than dark matters. Although complicated
physical models are needed to know the modified mass distribution, we
simply assume that the mass distribution of galaxies is well described
by the Singular Isothermal Sphere (SIS) approximation: 
\begin{equation}
\rho(r)=\frac{\sigma^2}{2\pi Gr^2},
\end{equation}
where $\sigma$ is a one-dimensional velocity dispersion and is related to
the circular velocity $v_{\rm c}$ as $v_{\rm c}=\sqrt{2}\sigma$. We also
assume that $v_{\rm c}$ as a function of halo mass $M$ can be
described as
\begin{equation}
 \gamma_v=\frac{v_{\rm c}}{v_{\rm vir}(M)},\label{gammav}
\end{equation}
where $v_{\rm vir}(M)$ is the halo virial velocity defined by $v_{\rm
vir}(M)=\sqrt{GM/r_{\rm vir}}$ and $\gamma_v$ is an arbitrary constant.
The SIS 
approximation for galaxies is consistent with several observations such as
dynamics \citep[e.g.,][]{rix97} and gravitational lensing 
\citep[e.g.,][]{rusin01a,cohn01}.

\subsection{Velocity Function of Galaxies\label{sec:th:vf}}
Once the probability $p_g(M)$ and the galaxy circular velocity $v_{\rm
c}(M)$ are given, we can calculate the velocity function  of galaxies
from the mass function of dark halos: 
\begin{equation}
\frac{dn}{dv_{\rm c}}(v_{\rm c}(M))=p_{\rm g}(M)\left|\frac{dv_{\rm c}(M)}{dM}\right|^{-1}\frac{dn}{dM}(M).
\label{vf}
\end{equation}
For the mass function $dn/dM$, we adopt a fitting form derived by
\citet{sheth99}: 
\begin{eqnarray}
\frac{dn_{\rm ST}}{dM}&=&0.322\left[1+\left(\frac{\sigma_M^2}
{0.707\delta_0^2(z)}\right)^{0.3}\right]\sqrt{\frac{1.414}{\pi}}\nonumber\\
&&\times\frac{\rho_0}{M}\frac{\delta_0(z)}{\sigma_M^2}\left|
\frac{d\sigma_M}{dM}\right
|\exp\left[-\frac{0.707\delta_0^2(z)}{2\sigma_M^2}\right].
\label{st}
\end{eqnarray}
This fitting form coincides more accurately with numerical
simulations than the analytic mass function derived by \citet{press74}.
The velocity function is sometimes expressed in terms of $\log v_{\rm
c}$ as
\begin{equation}
 \Psi(v_{\rm c})=\frac{dn}{d\log v_{\rm c}}.
\end{equation}
In this calculation we neglect the contribution from ``substructures'' 
(i.e., galaxies in groups and clusters), because this mainly changes the 
normalization of the velocity function which we do not use as a
constraint on the model parameters (see \S \ref{sec:const:vf}).
Substructures, however, may affect the fraction of dark halo lenses
directly, because it changes only the number of galaxy lenses. Therefore
we consider the effect of substructures in predicting the fraction of
dark halo lenses (see \S \ref{sec:frac}). 

\subsection{Lensing Probability Distribution}
Bearing the picture described in \S \ref{sec:th:baryon} in mind, we calculate 
the probability of gravitational lensing caused by bright galaxies and
dark halos separately: 
\begin{eqnarray}
P_{\rm gal}(>\theta; z_{\rm S}, L)=&&\nonumber\\
&&\hspace{-15mm}\int_0^{z_{\rm S}}dz_{\rm L}\int_{M_{\rm min}}^{\infty} dM\,p_{\rm g}(M)\sigma_{\rm SIS}\,B\frac{c\,dt}{dz_{\rm L}}(1+z_{\rm L})^3\frac{dn}{dM},\label{pgal}\\
P_{\rm dark}(>\theta; z_{\rm S}, L)=&&\nonumber\\
&&\hspace{-25mm}\int_0^{z_{\rm S}}dz_{\rm L}\int_{M_{\rm min}}^{\infty} dM\,\left\{1-p_{\rm g}(M)\right\}\sigma_{\rm NFW}\,B\frac{c\,dt}{dz_{\rm L}}(1+z_{\rm L})^3\frac{dn}{dM},\label{pdark}
\end{eqnarray}
where $z_{\rm S}$ is the source redshift, $z_{\rm L}$ is the lens redshift, 
$L$ is the luminosity of the source, and $dn/dM$ is the comoving number 
density of dark halos (eq. [\ref{st}]). 
Lensing cross sections $\sigma_{\rm SIS}$ and $\sigma_{\rm NFW}$ are given by
\begin{eqnarray}
 \sigma_{\rm SIS}&=&16\pi^3\left(\frac{\sigma}{c}\right)^4\left(\frac{D_{\rm OL}D_{\rm LS}}{D_{\rm OS}}\right)^2,\label{cs_sis}\\
 \sigma_{\rm NFW}&=&\pi\left(\eta_{\rm rad}\frac{D_{\rm OL}}{D_{\rm OS}}\right)^2,
\end{eqnarray}
where $\eta_{\rm rad}$ is the radius of the radial caustic at source
plane \citep*[e.g.,][]{schneider92} and $D_{\rm OL}$, $D_{\rm OS}$, 
and $D_{\rm LS}$ are the angular diameter distances to the lens, to the
source, and between the lens and source, respectively. The
one-dimensional velocity dispersion $\sigma$ in equation (\ref{cs_sis})
can be represented as a function of halo mass $M$:
\begin{equation}
 \sigma=\frac{v_{\rm c}}{\sqrt{2}}=\frac{\gamma_vv_{\rm vir}(M)}{\sqrt{2}},
\end{equation}
where equation (\ref{gammav}) is used. The lower limit of integral by
mass, $M_{\rm min}$, is determined by solving the equation
$\theta=\theta(M_{\rm min},z_{\rm S},z_{\rm L})$. The magnification bias
\citep{turner80,turner84} is included in $B$ as follows:
\begin{equation}
B=\frac{1}{\sigma_{\rm lens}\Phi(z_{\rm S}, L)}\int_{\rm multi}d^2\eta\,\Phi(z_{\rm S}, L/\mu(\vec{\eta}))\frac{1}{\mu(\vec{\eta})},
\label{biasfactor}
\end{equation}
where $\Phi(z_{\rm S}, L)$ is the luminosity function of sources and 
$\mu(\vec{\eta})$ is the magnification factor at $\vec{\eta}$.

The total lensing probability with image separation larger than $\theta$
is given by 
\begin{equation}
P(>\theta; z_{\rm S}, L)=P_{\rm gal}(>\theta; z_{\rm S}, L)+P_{\rm dark}(>\theta; z_{\rm S}, L).
\label{dist}
\end{equation}

%%%%%%%%%%%%%%%%%%%%%%%%%%%%%%%%%%%%%%%%%%%%%%%%%%%%%%%%%%%
%%%%%%%%%%%%%%%%%%%%%%%%%%%%%%%%%%%%%%%%%%%%%%%%%%%%%%%%%%%
\section{Constraints on the Model\label{sec:const}}
%%%%%%%%%%%%%%%%%%%%%%%%%%%%%%%%%%%%%%%%%%%%%%%%%%%%%%%%%%%
%%%%%%%%%%%%%%%%%%%%%%%%%%%%%%%%%%%%%%%%%%%%%%%%%%%%%%%%%%%
\subsection{Velocity Function of Galaxies\label{sec:const:vf}}
The velocity function is valuable because it is easy to handle
theoretically, and is useful to test the model of galaxy formation and
cosmology  \citep{cole89,shimasaku93,gonzalez00,kochanek01b}.
Using the model described in \S \ref{sec:th:vf}, we can calculate the
velocity function of galaxies from the mass function of dark halos. 
By comparing the theoretical velocity function with observed velocity
functions, one can constrain the model parameters such as $\delta_{\rm
h}$, $M_{\rm h}$, and $\gamma_v$. The definite theoretical prediction,
however, needs the {\it correct} value of cosmological parameters such
as $\sigma_8$. Therefore we use the normalized velocity function 
\begin{equation}
 \psi(v_{\rm c})\equiv\frac{\Psi(v_{\rm c})}{\Psi(200{\rm km/s})},
 \label{nvf}
\end{equation}
instead of $\Psi(v_{\rm c})$. Moreover, by using $\psi(v_{\rm c})$ we
can neglect the effect of the overall factor of $p_{\rm g}(M)$. 
For the observed velocity functions, we use five velocity functions
derived by \citet{gonzalez00}: velocity functions derived from Southern
Sky Redshift Survey (SSRS2), Automatic Plate Measuring facility survey
(APM), United Kingdom Schmidt Telescope survey (UKST), Las Campanas
Redshift Survey (LCRS), and K-band survey by \citet{gardner97}. We
restrict the comparison with observations in the region $200{\rm
km/s}<v_{\rm c}<500{\rm km/s}$ because of the following two reasons. One
reason is that observed velocity functions are not reliable at the
high velocity region ($v_{\rm c}>500{\rm km/s}$) \citep{gonzalez00}. The
other reason is the obvious difference between observed and theoretical number
density of galaxies at the low velocity region ($v_{\rm c}<200{\rm
km/s}$). The reason of this difference is that in our model we neglect
effects of e.g., supernovae feedback \citep{dekel86} which substantially
suppress the number of galaxies in low-mass halos
\citep[e.g.,][]{gonzalez00,nagamine01}. However we are interested in the
gravitational lensing with angular separation larger than current
angular resolution ($\theta\sim0.3''$), thus we can safely neglect such
low-velocity galaxies which have little impact on our lensing results.

\begin{figure*}[t]
 \epsscale{1.2}
\plotone{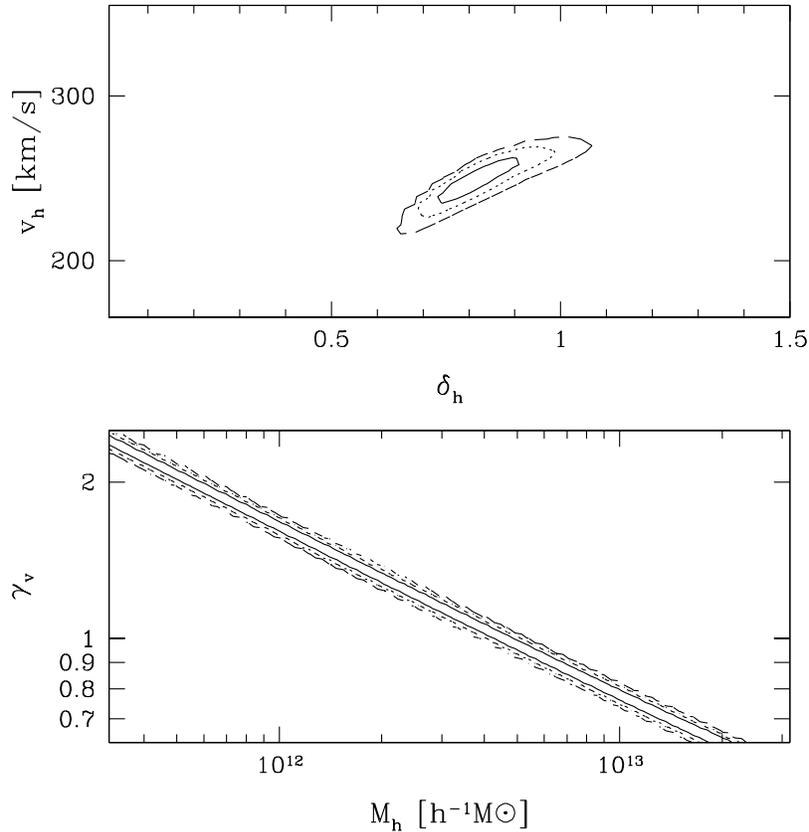} 
\caption{Constraints on model parameters from the velocity function of
 galaxies. 68\% ({\it solid}), 95\% ({\it dotted}), and 99\% ({\it
 dashed}) likelihood contours are shown. Upper panel: constraint
 on $\delta_{\rm h}$ and $v_{\rm h}\equiv v_{\rm c}(M_{\rm h})$. Lower panel:
 constraint on $M_{\rm h}$ and $\gamma_v$ assuming $\delta_{\rm h}=0.78$.   
\label{fig:chis}}  
\end{figure*}

Figure \ref{fig:chis} shows constraints on model parameters from 
observed velocity functions. Contours are calculated from ratios of 
combined likelihood ${\cal L}\propto\prod\exp(-\frac{1}{2}\chi_i^2)$.
Errors of observed velocity functions are estimated from errors of
fitting parameters of velocity functions \citep[see][]{gonzalez00}. As
easily seen from the upper panel of Figure \ref{fig:chis}, we can put
useful constraints on $\delta_{\rm h}$ and $v_{\rm h}\equiv v_{\rm
c}(M_{\rm h})$. The best fit parameter set is $(\delta_{\rm h},v_{\rm
h})\simeq(0.78,238{\rm km/s})$. This means that there still remains 
strong degeneracy between $M_{\rm h}$ and $\gamma_v$: parameter
combinations which yield the same $v_{\rm h}$ cannot be discriminated
from the velocity functions. This fact is also shown in the lower
panel of Figure \ref{fig:chis}. One of our best fit parameter sets 
$(\delta_{\rm h},M_{\rm h},\gamma_v)=(0.78,10^{12}h^{-1}M_\odot,1.67)$
is compared with observed velocity functions in Figure \ref{fig:vf}. It
is clearly shown that the prediction of our model shows good coincidence
with observed velocity functions. 

\vspace{0.5cm}
\centerline{{\vbox{\epsfxsize=8.3cm\epsfbox{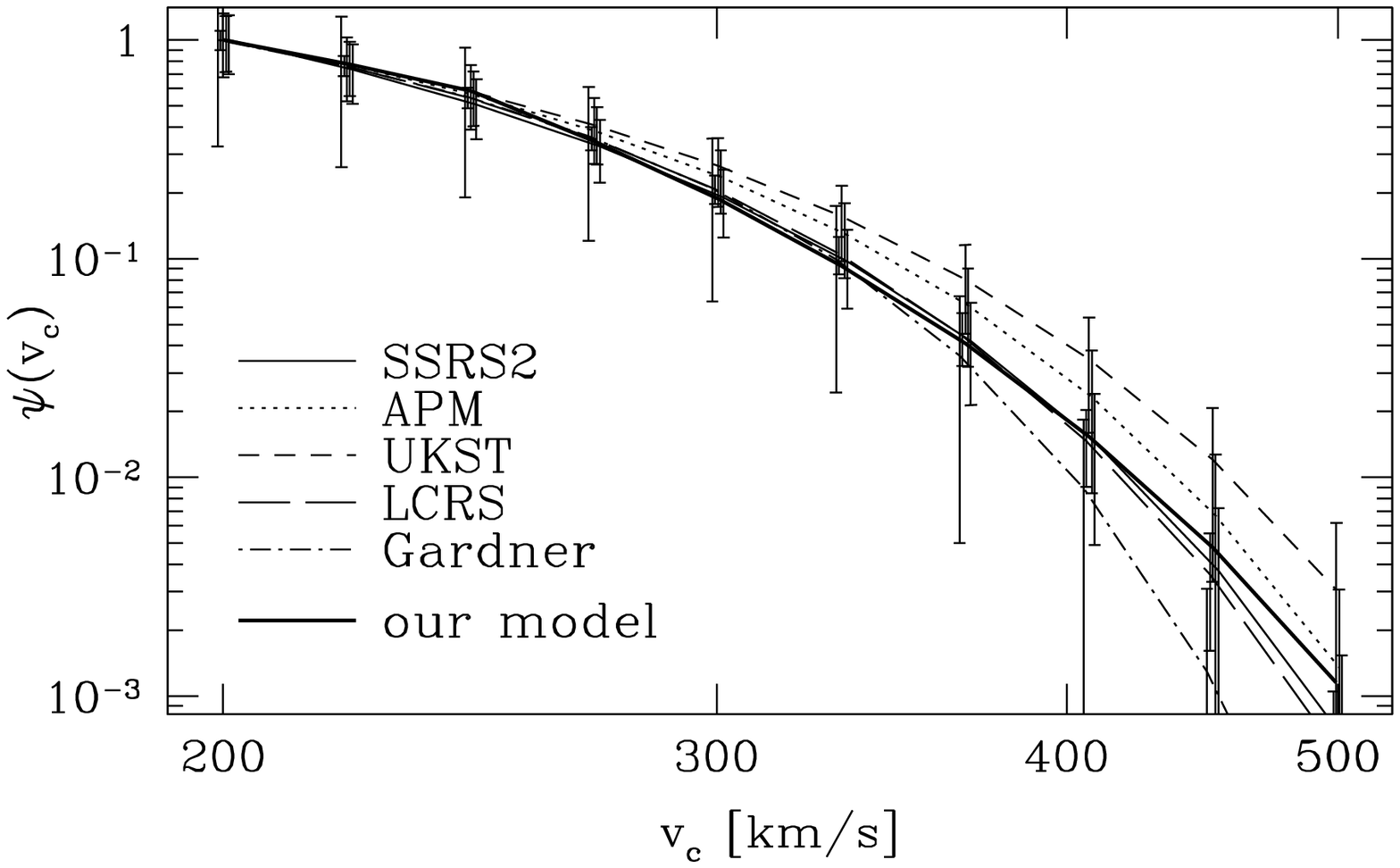}}}}
\figcaption{One of our best fit models for the velocity function is compared 
with the five observed velocity functions \citep{gonzalez00}. Here
 normalized velocity functions (eq. [\ref{nvf}]) are compared. Errorbars
 are estimated from the error of fitting 
parameters of velocity functions. The model 
parameters are chosen so as to satisfy constrains in Figure \ref{fig:chis}: 
$\delta_{\rm h}=0.78$, $M_{\rm h}=10^{12}h^{-1}M_\odot$, and 
$\gamma_v=1.67$.\label{fig:vf}}
\vspace{0.5cm}

\subsection{Image Separation Distribution\label{sec:const:sep}}

The distribution of image separations in strong gravitational lensing
becomes another test of our model. The distribution is calculated from 
equation (\ref{dist}). For the observed lens sample, we consider the
Cosmic Lens All-Sky Survey \citep[CLASS;][]{helbig00}. This sample is
complete at image separations $0.3''<\theta<15''$
\citep{helbig00,phillips01}, and has 18 lenses among $\sim 12,000$ radio
sources. Sources have the flux distribution $dn/dS\propto S^{-2.1}$
\citep{rusin01b}. The mean redshift is estimated to be $\langle z_{\rm S}
\rangle=1.27$, although the redshift distribution of sources is still poorly
understood \citep{marlow00}.
As in the case of the velocity function, actually we consider the
normalized image separation distribution:
\begin{equation}
 p(>\theta; z_{\rm S}, L)=\frac{P(>\theta; z_{\rm S}, L)}{P(>0.3''; z_{\rm S}, L)},
\label{ndist}
\end{equation}
instead of usual probability distribution. This is because the absolute
probability may suffer from uncertainties of source redshifts,
magnification bias, and cosmological parameters. On the other hand the
distribution $p(>\theta)$ mainly contains information on the shape
of the mass function and effects of baryonic compression
\citep{kochanek01a,kochanek01b,keeton01a}. And the distribution
$p(>\theta)$ is quite insensitive to the source population and
cosmological parameters, thus can be used for samples in which the source
population is unknown such as CASTLES sample.

\begin{figure*}[t]
 \epsscale{1.5}
\plotone{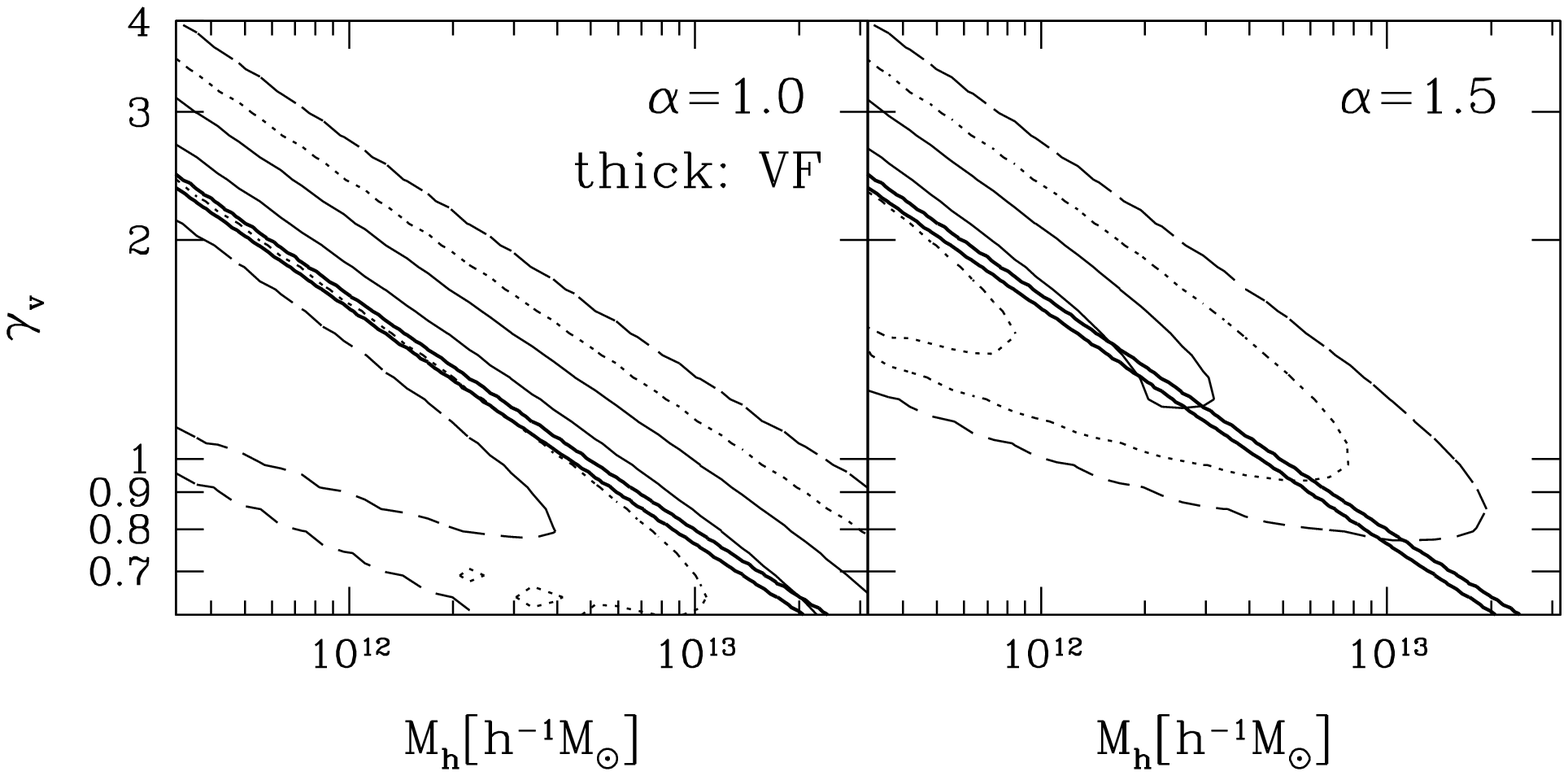} 
\caption{Constraints on model parameters from the distribution of image
 separations in strong gravitational lensing. Here we have assumed
 $\delta_{\rm h}=0.78$. 68\% ({\it solid}), 95\%
 ({\it dotted}), and 99\% ({\it dashed}) contours are calculated using
 Kolmogorov-Smirnov test. 68\% constraint from the velocity function (see Fig. 
 \ref{fig:chis}) is also plotted by thick lines. 
 For the inner slope of the density profile of dark
 halos (eq. [\ref{nfw}]), we consider both $\alpha=1.0$ ({\it left}) and
 $\alpha=1.5$ ({\it right}).
\label{fig:ks}}  
\end{figure*}

The distribution of image separations is tested by using a
Kolmogorov-Smirnov (KS) test. For the observed distribution, we use the
distribution of CLASS survey. The result is shown in Figure \ref{fig:ks}. We
plot contours in $M_{\rm h}$-$\gamma_v$ plane by fixing $\delta_{\rm
h}=0.78$ which is constrained from the velocity function (see Figure
\ref{fig:chis}). From this figure, we find that the constraint from the
distribution of image separations is consistent with the constraint from
the velocity function. The exception is the case of $\alpha=1.5$, where
$\alpha$ is the inner slope of the density profile of dark halos (eq.
[\ref{nfw}]). In this case, the lower value of $\gamma_v$ becomes
inconsistent with the observation, because the lensing cross section for
SIS is $\sigma_{\rm SIS}\propto \gamma_v^4$ (see eq. [\ref{cs_sis}]) and
the lower value of $\gamma_v$ means that the contribution of dark halo
lenses becomes more significant. That is, when $\gamma_v$ is low, the number
of dark halo lenses is comparable to that of galaxy lenses, thus a steep
cutoff in the distribution of image separations never appears.
Therefore, we conclude that the constraint from the distribution of
image separations is consistent with the constraint from the velocity
function, except for the cutoff of KS probability at low $\gamma_v$
which may appear when $\alpha$ is large. We also compare the
distribution of our model with the observed distribution in Figure
\ref{fig:sep}. The model parameters are chosen so as to satisfy the
constraint from the velocity function, and same as those used in Figure
\ref{fig:vf}. We also plot the distribution assuming that all lens
objects are well approximated by SIS. This assumption corresponds to
$p_{\rm g}(M)=1$ for all $M$. It is obvious that our model well
reproduces the observed distribution while the distribution of all SIS
assumption is far from the observed distribution, as pointed out 
by \citet{keeton98}.

\vspace{0.5cm}
\centerline{{\vbox{\epsfxsize=8.3cm\epsfbox{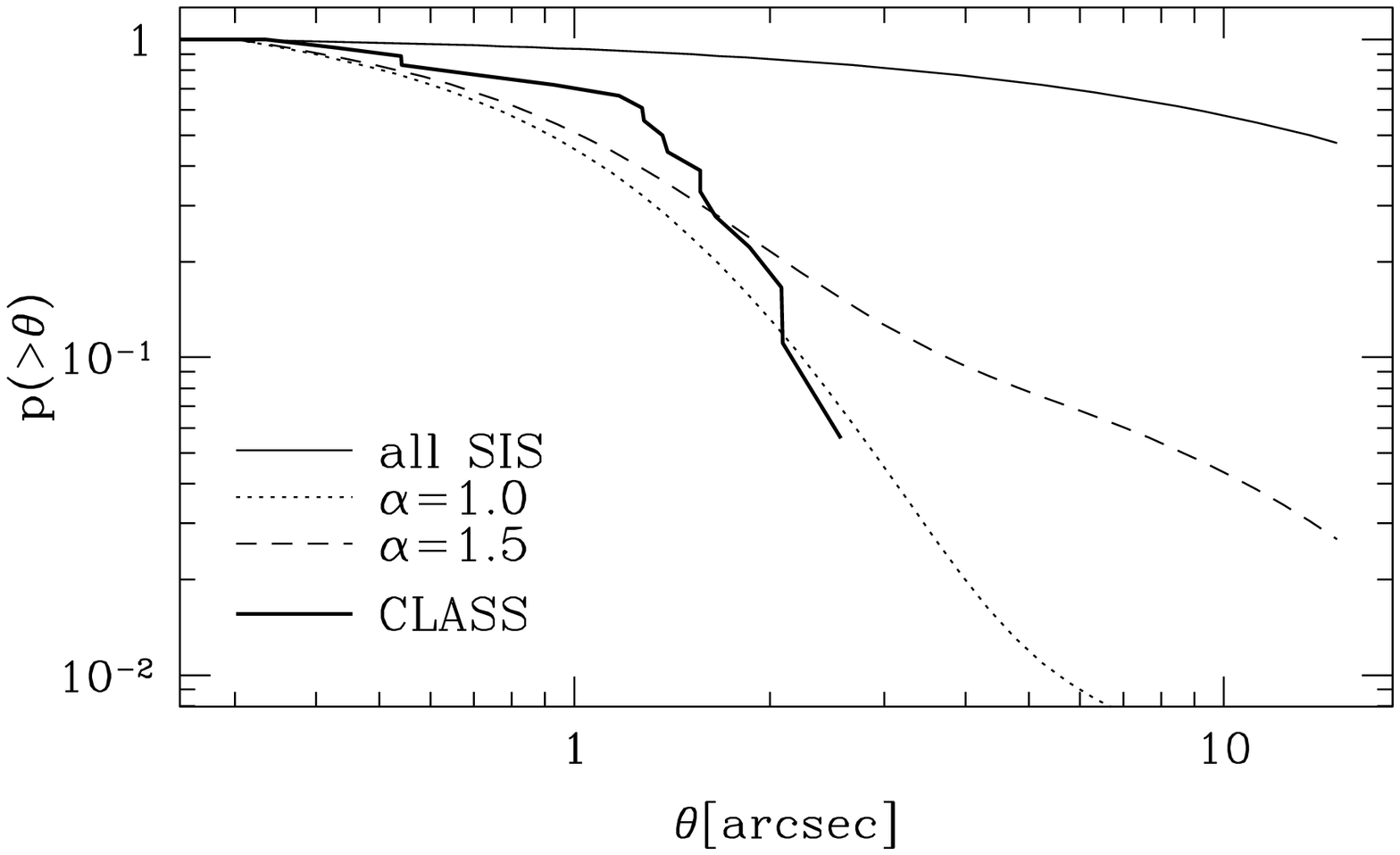}}}}
\figcaption{The distribution of image separations in gravitational lensing
 (eq. [\ref{ndist}]).
The model parameters are same as Figure \ref{fig:vf}. The cases with 
$\alpha=1.0$ ({\it dotted}) and $\alpha=1.5$ ({\it dashed}) are plotted. 
The observed 
distribution in CLASS survey is shown by thick solid line. The thin solid 
line indicates the theoretical distribution assuming all lens objects are 
SIS lenses. This corresponds to $p_{\rm g}(M)=1$ for all $M$.\label{fig:sep}}
\vspace{0.5cm}

\subsection{Various Constraints on the Circular Velocity of Galaxies\label{sec:const:vc}}
In \S \ref{sec:const:vf} and \S \ref{sec:const:sep}, we constrained
model parameters from the velocity function of galaxies and the
distribution of image separations. The strong degeneracy between $M_{\rm
h}$ and $\gamma_v$, however, still remains. Therefore in this subsection we
try to put constraints on $\gamma_v$ from various theories and observations.
We consider following three constraints. 

First, \citet{seljak02a} gave values of $\gamma_v$ at several halo
mass derived from the observation and analysis of galaxy-galaxy lensing
\citep{mckay02,guzik02} and Tully-Fisher/fundamental plane relations.
The observation of galaxy-galaxy lensing allows us to determine the mass
of dark halos and its relation to the luminosity of galaxies.
Tully-Fisher/fundamental plane relations are used to derive the galaxy
circular velocity from its luminosity. He found that $\gamma_v$ is
significantly larger than 1, $\gamma_v\sim1.8$ around $L_*$. He also
found the decrease of $\gamma_v$ from $L_*$ to $7L_*$, $\gamma_v\sim1.4$
at $7L_*$.  

Second, \citet{cole00} semi-analytically calculated average values of 
$\gamma_v$ for galaxies with $-20<M_I-5\log h<-18$. They calculated
$\gamma_v$ for various parameter sets of semi-analytic model, and found
that most of parameter sets predict $\gamma_v\sim1.3-1.4$. The most
important parameter for $\gamma_v$ is the baryon density $\Omega_{\rm
b}$, but even in the extreme cases they examined, $\Omega_{\rm b}=0.01$
and $0.04$, $\gamma_v$ becomes $1.13$ and $1.96$, respectively. 

Third, \citet{mo98} considered the analytic disk formation model on the
basis of adiabatic compression \citep[e.g.,][]{blumenthal86}. In their
model, $\gamma_v$ depends on the mass of cooled baryons and the angular
momentum of galactic disc. More specifically, their model is
characterized by following three parameters; the concentration parameter
$c$, the ratio of disk mass to halo mass $m_{\rm d}$, and the angular
momentum $\lambda'$ \citep[see][]{mo98}. We assume that the
concentration parameter is related to halo mass as in equation (\ref{bul}).
Then $\gamma_v$ at fixed mass depends on $m_{\rm d}$ and $\lambda'$. We
examine following two extreme cases: $(m_{\rm
d},\lambda')=(0.02,0.02)$ and $(0.02,0.2)$. Most of parameter
sets which are physically reasonable predict the value of $\gamma_v$
between these two extreme cases. 

\vspace{0.5cm}
\centerline{{\vbox{\epsfxsize=8.3cm\epsfbox{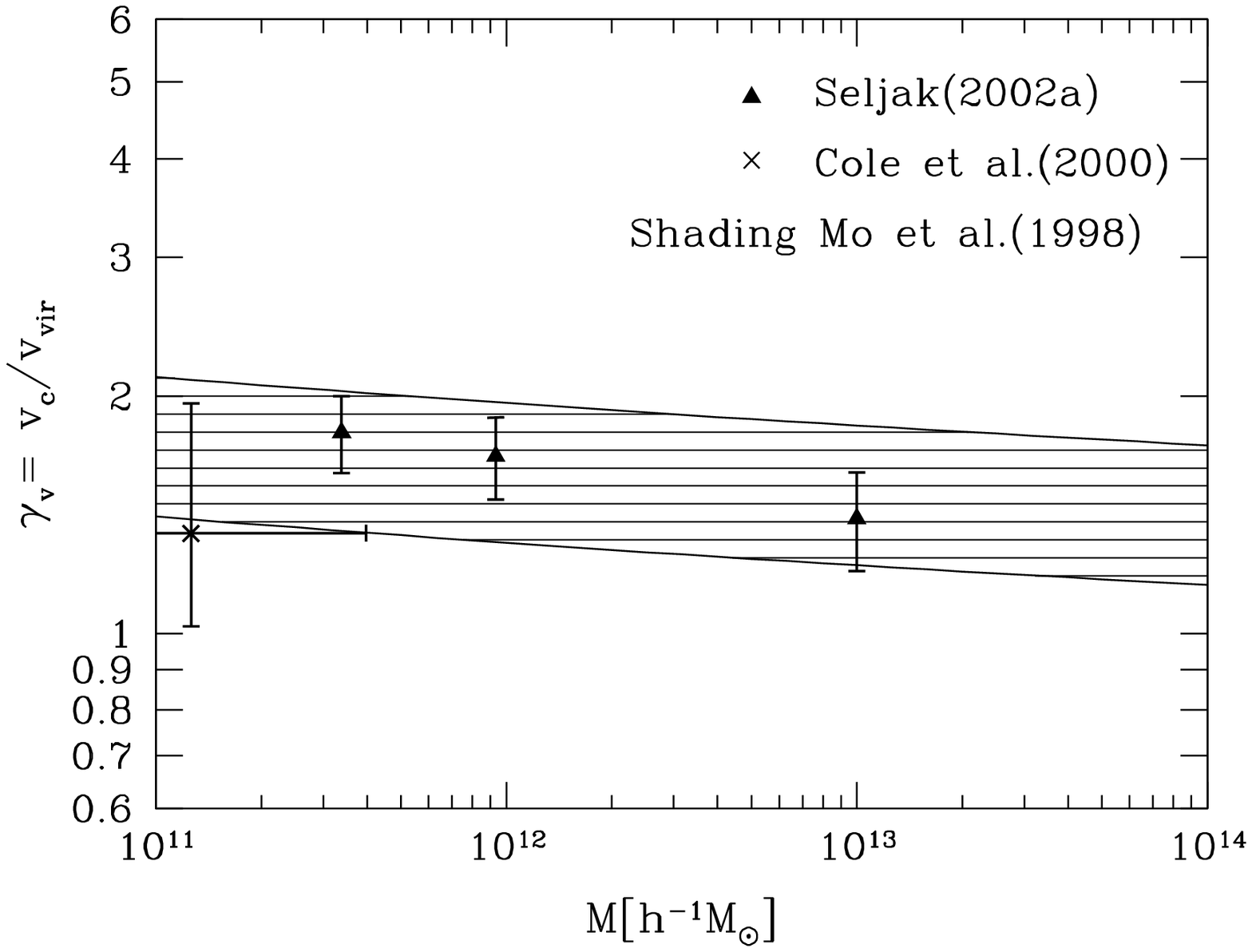}}}}
\figcaption{Constraints on $\gamma_v$ from various models. (i)
 \citet{seljak02a} gave values of $\gamma_v$ at several halo mass
 derived from the observation and  
analysis of galaxy-galaxy lensing \citep{mckay02,guzik02} and 
Tully-Fisher/fundamental plane relations. Errorbars indicate $1\sigma$ level 
error. (ii) \citet{cole00} semi-analytically calculated the average values of 
$\gamma_v$ for various parameter sets of semi-analytic model. The
 errorbar in the direction of $\gamma_v$-axis means that all parameter
 sets predict $\gamma_v$ within 
this range. (iii) In the analytic disk formation model of \citet{mo98}, 
$\gamma_v$ depends on the mass of cooled baryons and the angular momentum of 
galactic disc. Any choice of these two parameters which is physically allowed, 
however, results in the value of $\gamma_v$ in the shading region.
\label{fig:vcvvir}}
\vspace{0.5cm}

The result is summarized in Figure \ref{fig:vcvvir}. 
Particularly we are interested in the strong gravitational lensing, thus
we focus on the range $10^{12}h^{-1}M_\odot\lesssim M \lesssim
10^{13}h^{-1}M_\odot$ where gravitational lensing with separations
$\theta>0.3''$ is most efficient. In this range of halo mass, various
constraints indicate that $\gamma_v$ is restricted to $1\lesssim
\gamma_v\lesssim 2$. In particular, the upper bound of $\gamma_v$ is
quite robust because $\gamma_v$ is closely related to the amount of
cooled baryons; to produce $\gamma_v\gg 2$, we need the extraordinarily 
large amount of cooled baryon which exceeds the global baryon
mass ratio 
\citep{seljak02a}. Figure \ref{fig:vcvvir} also indicates that 
$\gamma_v$ should slightly depend on halo mass. The mass dependence of
$\gamma_v$, however, is not so important because our interest is
restricted in a narrow mass range. Therefore we can safely assume that
$\gamma_v$ is constant. 

\begin{figure*}[t]
 \epsscale{1.5}
\plotone{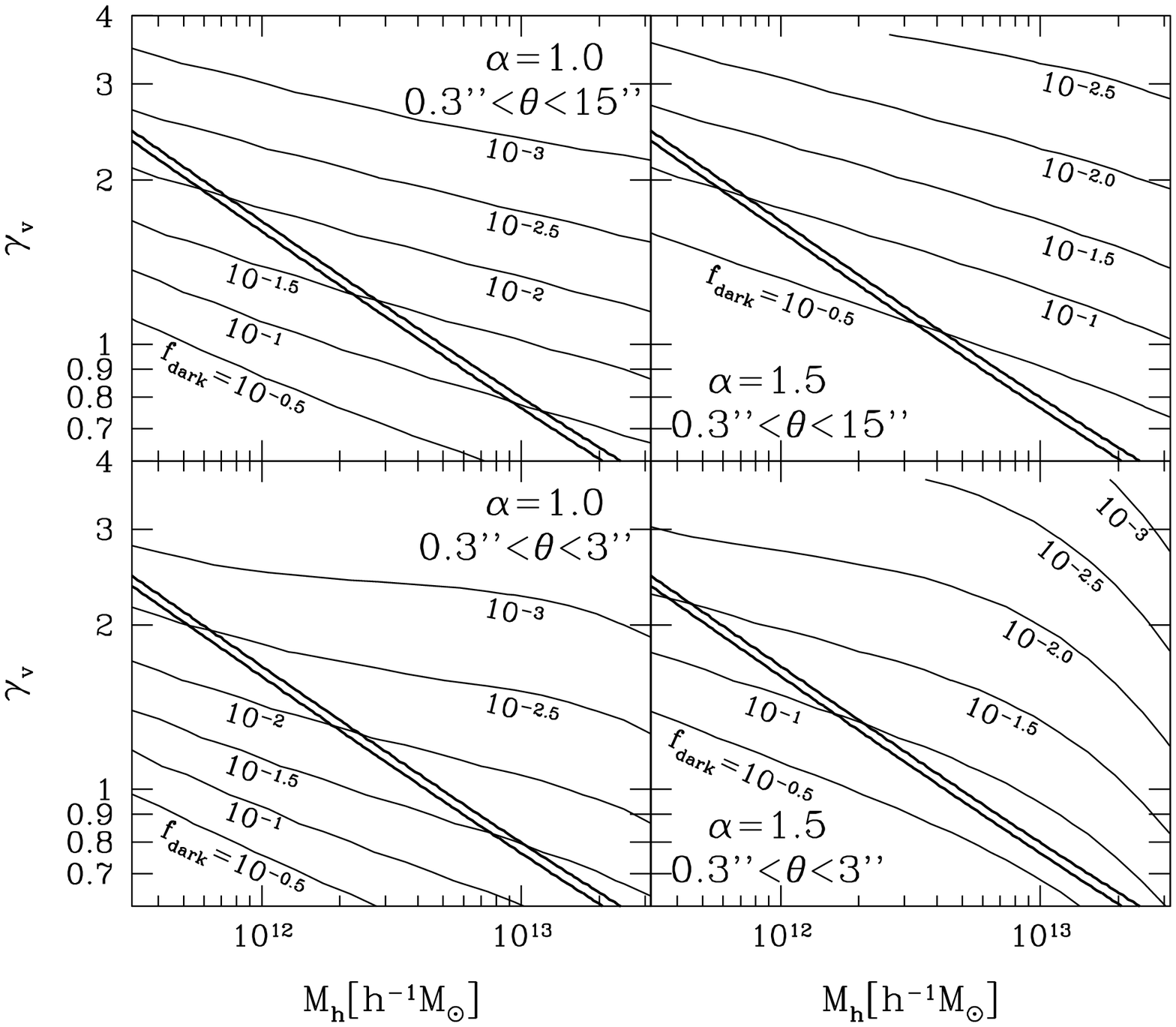} 
\caption{Contour plots for the fraction of dark halo lenses in $M_{\rm h}$-$\gamma_v$  
plane. Thick lines are 68\% contours derived from the velocity function 
(see Fig. \ref{fig:chis}). The inner cusps of dark halos (eq. [\ref{nfw}]) 
are $\alpha=1.0$ ({\it left}) and $\alpha=1.5$ ({\it right}). For the 
range of image separations, $0.3''<\theta<15''$ ({\it upper}) and 
$0.3''<\theta<3''$ ({\it lower}) are considered.}  
\label{fig:fd_cont}
\end{figure*}

%%%%%%%%%%%%%%%%%%%%%%%%%%%%%%%%%%%%%%%%%%%%%%%%%%%%%%%%%%%
%%%%%%%%%%%%%%%%%%%%%%%%%%%%%%%%%%%%%%%%%%%%%%%%%%%%%%%%%%%
\section{The Fraction of Dark Halo Lenses\label{sec:frac}}
%%%%%%%%%%%%%%%%%%%%%%%%%%%%%%%%%%%%%%%%%%%%%%%%%%%%%%%%%%%
%%%%%%%%%%%%%%%%%%%%%%%%%%%%%%%%%%%%%%%%%%%%%%%%%%%%%%%%%%%

According to the model described in \S \ref{sec:th}, we predict the
fraction of dark halo lenses. In our model, there are three
parameters which govern effects of baryonic compression;
$\delta_{\rm h}$, $M_{\rm h}$, and $\gamma_v$. Since the constraint from
the velocity function of galaxies (\S \ref{sec:const:vf}) indicates that
$\delta_{\rm h}$ should be restricted around $\delta_{\rm h}\sim 0.78$,
we fix $\delta_{\rm h}=0.78$ in the remainder of this paper. On the
other hand, the other
parameters, $M_{\rm h}$ and $\gamma_v$, were poorly constrained from the
velocity function of galaxies and the distribution of image separations.
Therefore we should see the
dependence of the fraction of dark halo lenses on both $M_{\rm h}$ and
$\gamma_v$. Another important parameter we should examine is the inner
slope of density profile ($\alpha$ in eq. [\ref{nfw}]), because it is
also known that the number of dark halo lenses is extremely sensitive to
$\alpha$ \citep{wyithe01,keeton01b,takahashi01,li02,oguri02a}. 

The fraction of dark halo lenses at image separations
$\theta_1<\theta<\theta_2$ is given by 
\begin{equation}
f_{\rm dark}(\theta_1<\theta<\theta_2)=\frac{P_{\rm dark}(>\theta_1)-P_{\rm dark}(>\theta_2)}{P(>\theta_1)-P(>\theta_2)},\label{fdark}
\end{equation}
where $P_{\rm dark}(>\theta)$ and $P(>\theta)$ are calculated from
equations (\ref{pdark}) and (\ref{dist}), respectively. Again, $f_{\rm
dark}$ is almost independent of the source population (source redshift
and flux distribution) and cosmological parameters because these
primarily change the normalization of the overall lensing rate which we
never use. For the range of image separations, we consider following two
cases: (1) $0.3''<\theta<15''$. In this range of image separations,
CLASS survey is complete \citep{helbig00,phillips01}. Among 18 lenses
observed in CLASS survey, virtually all lenses are known to be galaxy
lenses \citep{rusin02}. (2) $0.3''<\theta<3''$. Most of current
ambiguous quasar pairs which may be dark halo lenses have separations
$\theta>3''$ and almost all lenses with smaller separations are known to be
produced by normal 
galaxies \citep{kochanek99,rusin02}. Therefore, in this case we can use
all lenses with separations $0.3''<\theta<3''$ in comparison with our
result. Below we assume that there is no dark halo lens for both cases.

\begin{figure*}[t]
\epsscale{1.5}
\plotone{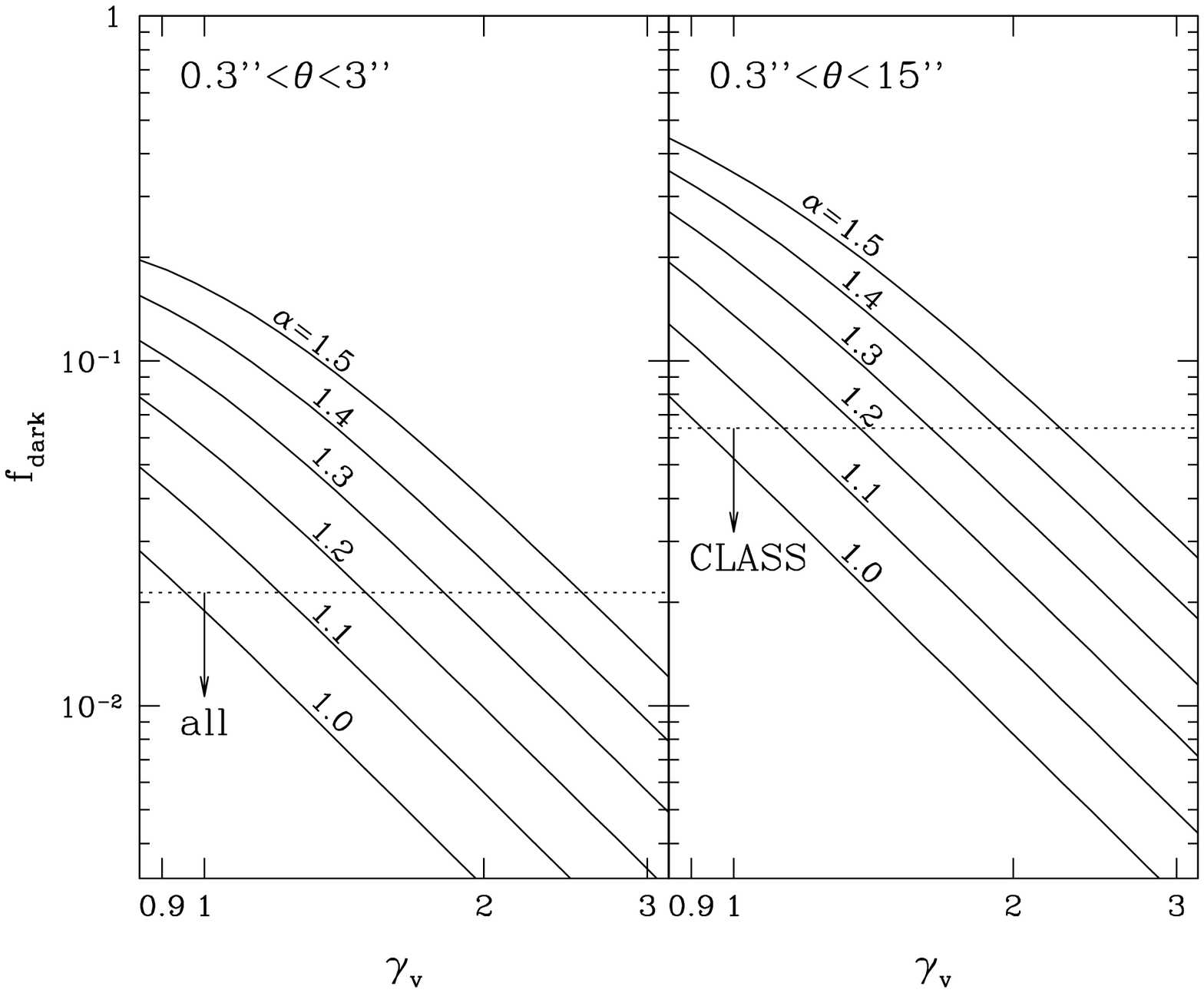} 
\caption{The fraction of dark halo lenses as a function of 
$\gamma_v$. Each solid line is the result with particular choice of $\alpha$. 
For each $\gamma_v$, the value of $M_{\rm h}$ is chosen so that it
 satisfies the constraint  
from the velocity function. Observational upper limits of $f_{\rm dark}$ at
 1$\sigma$ level (all lenses and the CLASS survey lenses) are also shown
 by dotted lines with downward arrows. } 
\label{fig:fd}
\end{figure*}

Figure \ref{fig:fd_cont} plots contours of $f_{\rm dark}$ in $M_{\rm
h}$-$\gamma_v$ plane. The constraint from the velocity function of
galaxies is also shown. From this figure, it is found that the fraction
of dark halo lenses is sensitive to $\gamma_v$: $f_{\rm dark}$ increases
as $\gamma_v$ decreases. The reason is that the cross section of SIS
lenses scales as $\sigma_{\rm SIS}\propto \gamma_v^4$ (see eq.
[\ref{cs_sis}]) and the number of galaxy lenses decreases as $\gamma_v$
decreases. The fraction of dark halo lenses also increases as $M_{\rm
h}$ decreases, because the number of dark halos which act as dark halo
lenses becomes large. Figure \ref{fig:fd_cont} indicates that $f_{\rm
dark}$ cannot be determined uniquely, even if we restrict $M_{\rm h}$
and $\gamma_v$ such that they satisfy the constraint from the velocity
function of galaxies. This fact is clearly shown in Figure \ref{fig:fd}.
In Figure \ref{fig:fd}, we plot $f_{\rm dark}$ as a function of
$\gamma_v$. The parameter $M_{\rm h}$ is chosen so that it satisfies
the constraint from the velocity function of galaxies. This figure
indicates that $f_{\rm dark}$ is sensitive to $\gamma_v$. Even if we
restrict $1\leq \gamma_v\leq 2$ as discussed in \S \ref{sec:const:vc},
$f_{\rm dark}$ has uncertainty of about one order of magnitude. Figure
\ref{fig:fd} also shows that $f_{\rm dark}$ is sensitive to $\alpha$.
From $\alpha=1.0$ to $1.5$, $f_{\rm dark}$ also changes about one order
of magnitude. Therefore, our conclusion is that robust predictions of
$f_{\rm dark}$ need information on both $\gamma_v$ and $\alpha$. If
the density profile of dark halos is determined from several
observations, we can break the degeneracy between $M_{\rm h}$
and $\gamma_v$ from the observation of $f_{\rm dark}$.

\begin{figure*}[t]
\epsscale{1.5}
\plotone{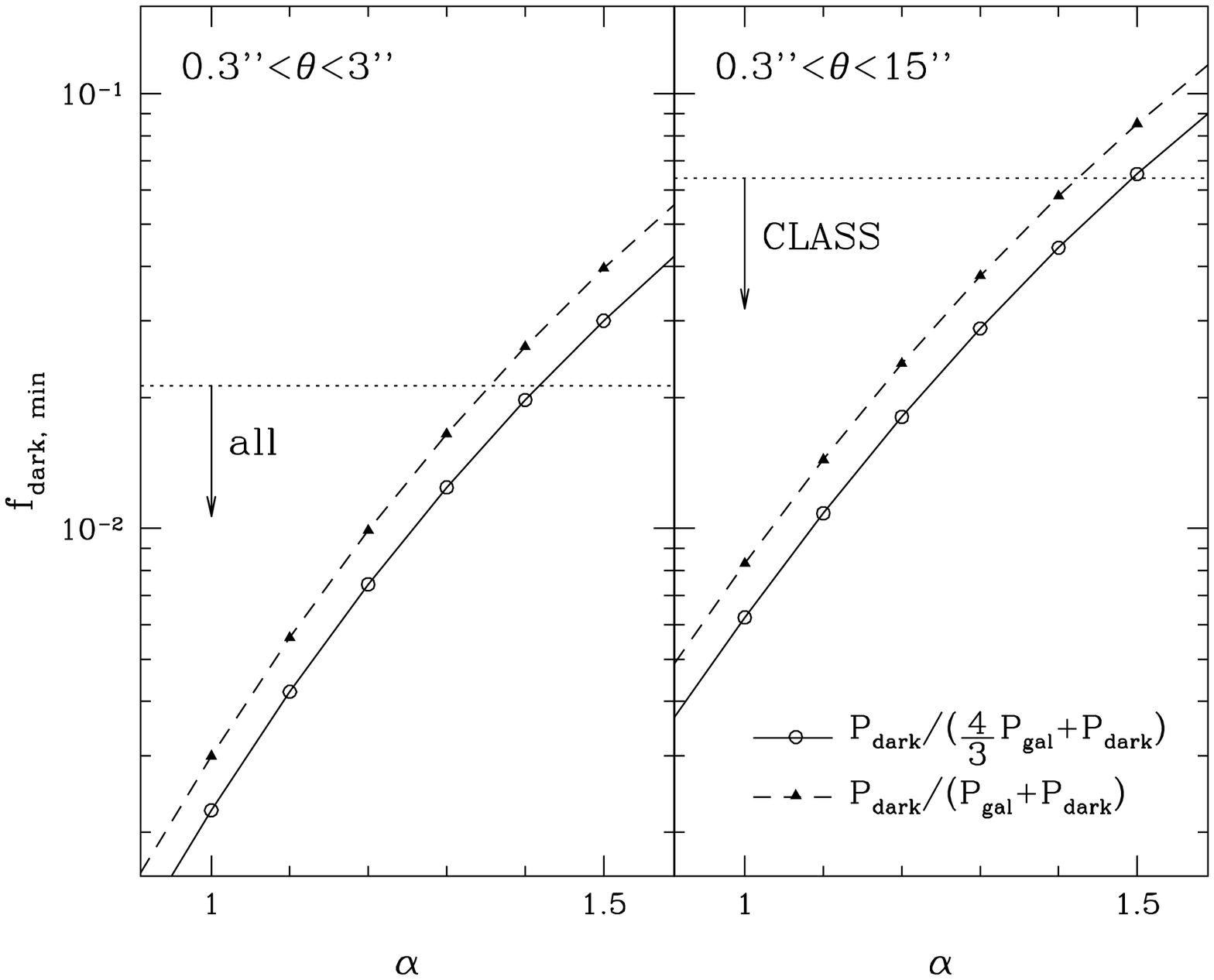} 
\caption{The lower limit of $f_{\rm dark}$ as a function of inner slope 
$\alpha$, when $\gamma_v$ is restricted to $\gamma_v\leq 2$. Dashed lines are 
calculated simply by setting $\gamma_v=2$ in Figure \ref{fig:fd}. For solid lines, 
the effect of galaxies which lie in groups/clusters is taken into account 
(see text for detail). Dotted lines  with downward arrows indicate the
 1$\sigma$ constraint from the lack of dark halo lenses in observations.}  
\label{fig:fd_alpha}
\end{figure*}

Although $f_{\rm dark}$ strongly depends on $\gamma_v$, we obtain the lower
limit $f_{\rm dark,min}$ by restricting $\gamma_v$. Discussions in \S
\ref{sec:const:vc} suggest that it is safe to adopt $\gamma_v\leq 2$. Thus by
setting $\gamma_v=2$ we can derive $f_{\rm dark,min}$, which is shown in
Figure \ref{fig:fd_alpha}. In this figure $f_{\rm dark, min}$ is plotted
against the inner slope $\alpha$. The lack of dark halo lenses places upper
limit on the value of $f_{\rm dark}$ which is also shown in Figure
\ref{fig:fd_alpha}. From this figure, we find that too cuspy inner slope
of dark halos ($\alpha\gtrsim 1.5$) is inconsistent with the lack of
dark halo lenses: even if we adopt sufficiently large $\gamma_v$,
$\gamma_v=2$, we predict too much dark halo lenses to reconcile with the
observation. This constraint has somewhat different meaning from the one
derived from the statistics of wide separation lensing
\citep{keeton01b,li02} because such statistics may suffer from the
uncertainties of cosmological parameters and the source population.

Are there any other effects which may change $f_{\rm dark}$? We consider
following two effects which may change our results. One is  
the effect of ``substructures'' (i.e., galaxies in groups and
clusters) as discussed in \S \ref{sec:th:vf}. Substructures may affect
the fraction of dark halo lenses directly, because it changes only the number
of galaxy lenses. Moreover the situation that lens galaxies which lie in
groups or 
clusters is not rare (e.g., Q0957+561), thus the effect should be
addressed quantitatively. Although the effect of substructures seems
difficult to estimate, \citet*{keeton00} calculated the fraction of
lensing galaxies which lie in groups and clusters. They predicted that
$\sim 25\%$ of lens galaxies are likely to be in groups or clusters.
Following this, the effect of substructures are simply estimated by
replace the probability of galaxy lensing as $P_{\rm gal}\rightarrow
(4/3)P_{\rm gal}$. This decreases $f_{\rm dark}$ by a factor $3/4$ at
most. The fraction of dark halo lenses including this effect is also
displayed in Figure \ref{fig:fd_alpha}. As seen in the figure, our
result that $\alpha\gtrsim 1.5$ is inconsistent with the observation
is not so affected by the effect of substructures. The other effect is
the existence of ``empty halos'', that is, dark halos which do not host
a galaxy. 
We have adopted the galaxy formation probability $p_{\rm g}(M)$ of the form
(\ref{gfp}), and this means that we have assumed there is no empty halo
at $M<M_{\rm h}$. Although this assumption does not conflict with the
observation of galaxy-galaxy lensing \citep{seljak02b}, it is still possible
that the significant amount of empty halos exists at galaxy-mass scale.
This effect, however, only {\it increases} $f_{\rm dark}$. Therefore {\it
lower limit} of $f_{\rm dark}$ used in Figure \ref{fig:fd_alpha} is
never changed by this effect. 

\begin{figure*}[t]
\epsscale{1.5}
\plotone{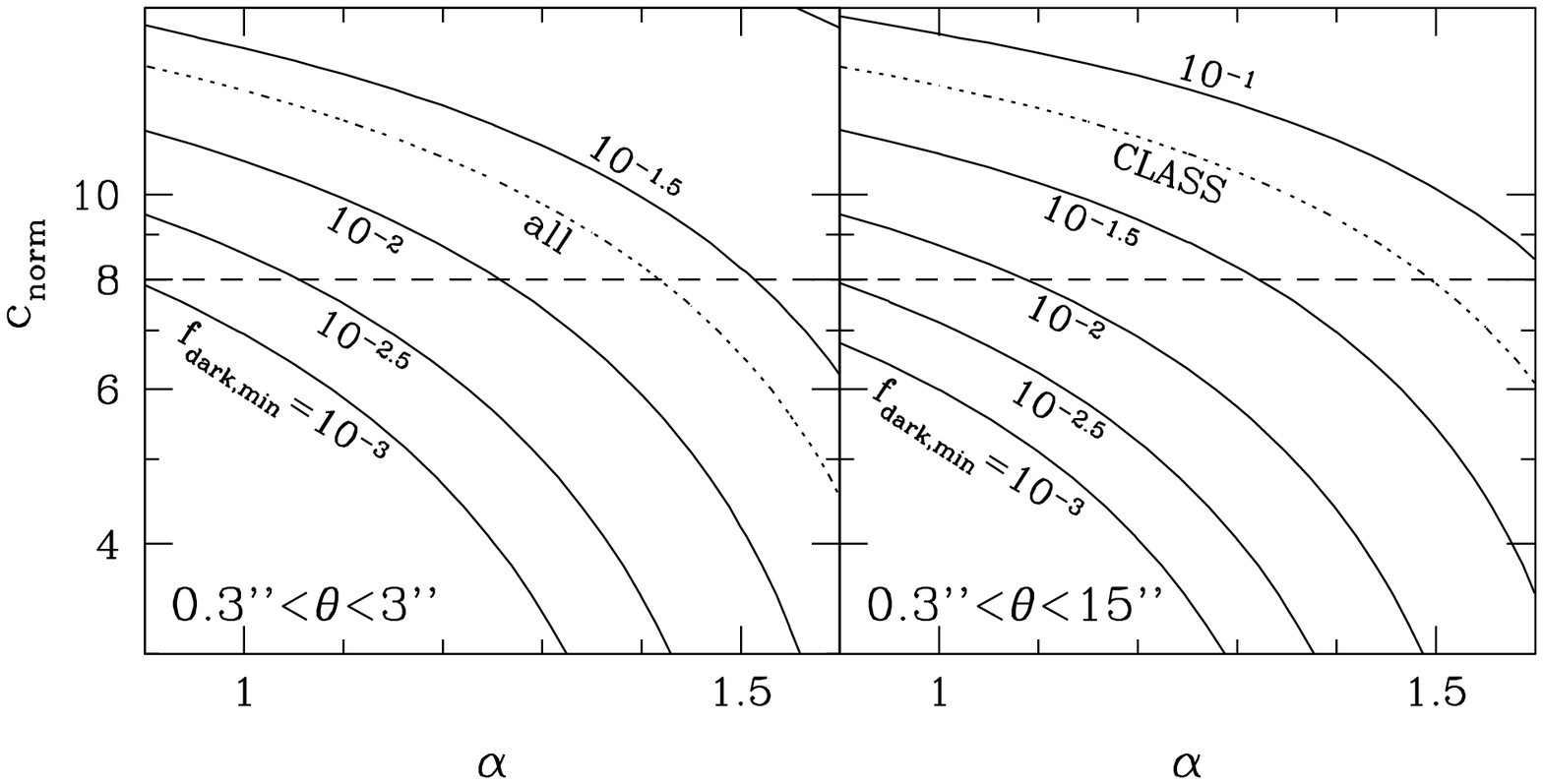} 
\caption{Contour plots for the lower limit of $f_{\rm dark}$ in the
 $\alpha$-$c_{\rm norm}$ plane. The effect of galaxies in
 groups/clusters is taken into account. Dotted lines show the 1$\sigma$
 constraint from the lack of dark halo lenses in observations. Dashed
 lines indicate the value of $c_{\rm norm}$ we used in the previous
 plots. }  
\label{fig:fdmin_cont}
\end{figure*}

The main result of \citet{wyithe01} and \citet{keeton01b} is that the
lensing probability of dark halos depends strongly on the choice of 
concentration parameters as well as inner slopes $\alpha$. Therefore, in
Figure \ref{fig:fdmin_cont} we examine $f_{\rm dark,min}$ for different
concentration parameters. In calculating this, we adopt the concentration
parameter of the following form: 
\begin{equation}
 c_{\rm vir}(M, z)=\frac{c_{\rm norm}}{1+z}\left(\frac{M}{10^{14}h^{-1}M_{\odot}}\right)^{-0.13},\label{bul2}
\end{equation}
instead of equation (\ref{bul}), and regard $c_{\rm norm}$ as a free
parameter. We plot $f_{\rm dark,min}$ in $\alpha$-$c_{\rm norm}$ plane.
From this figure, it is clearly seen that there is a strong degeneracy
between $\alpha$ and $c_{\rm norm}$ as reported by \citet{wyithe01} and
\citet{keeton01b}. Thus actually we can constrain the combination of
$\alpha$ and $c_{\rm norm}$, or the core mass fraction proposed by
\citet{keeton01b}, instead of $\alpha$ and $c_{\rm norm}$ separately.

%%%%%%%%%%%%%%%%%%%%%%%%%%%%%%%%%%%%%%%%%%%%%%%%%%%%%%%%%%%
%%%%%%%%%%%%%%%%%%%%%%%%%%%%%%%%%%%%%%%%%%%%%%%%%%%%%%%%%%%
\section{Conclusion\label{sec:conc}}
%%%%%%%%%%%%%%%%%%%%%%%%%%%%%%%%%%%%%%%%%%%%%%%%%%%%%%%%%%%
%%%%%%%%%%%%%%%%%%%%%%%%%%%%%%%%%%%%%%%%%%%%%%%%%%%%%%%%%%%
We have studied the effect of baryonic compression assuming the simple
parametric model used by \citet{kochanek01a}. Our model has following two
elements: the galaxy formation probability $p_{\rm g}(M)$ (eq.
[\ref{gfp}]) which describes the global efficiency of baryonic
compression, and the ratio of circular velocities of galaxies to
virial velocities of dark halos $\gamma_v=v_{\rm c}/v_{\rm vir}$ which
means how the inner structure of dark halos is modified due to
baryonic compression.  The model parameters are constrained from the
observed velocity function of galaxies and the distribution of image
separations in strong gravitational lensing, although the strong
degeneracy between model parameters still  remains. By using this model,
we predict the fraction of dark halo lenses $f_{\rm dark}$ (eq.
[\ref{fdark}]). Here  dark halo lenses mean the lens systems which are
produce by the gravitational potential of dark halos. The fraction of
dark halo lenses is independent of the normalization of total lensing
rate, thus is insensitive to cosmological parameters and information
of sources such as redshift and flux distribution. Instead the fraction
of dark halo lenses is expected to have information on both the
effect of baryonic compression and the density profile of dark halos
such as the inner density profile for which we modeled $\rho\propto
r^{-\alpha}$. We found that $f_{\rm dark}$ is indeed sensitive to the
inner slope $\alpha$, concentration parameter $c_{\rm norm}$, and
model parameters such as $\gamma_v$. Therefore,
definite predictions of $f_{\rm dark}$ need correct knowledge of
baryonic compression as well as the density profile of dark halos. This
also means that we can constrain the model of baryonic compression from
the observation of $f_{\rm dark}$ if the density profile of dark halos
is well known. 

Although the fraction of dark halo lenses is difficult to predict, we
can still derive the lower limit of $f_{\rm dark}$ by restricting
$\gamma_v\leq2$ which is inferred from various theories and observations
(see \S \ref{sec:const:vc}). We found that the steep inner profiles
($\alpha\gtrsim1.5$) or too centrally concentrated dark halos are
inconsistent with the lack of dark halo lenses in observations. As
described above, our result is quite insensitive to cosmological
parameters and source population. Therefore, our result is complementary
to the result of \citet{keeton01b} who obtained similar constraint on
the density profile of dark halos using the total lensing probability
which also depends strongly on cosmological parameters. 
One of possible systematic effects which may change $f_{\rm dark}$ is
the effect of galaxies in groups or clusters. By using the result of
\citet{keeton00}, we found that this effect changes $f_{\rm dark}$ by a
factor $3/4$ at most. Therefore this effect does not change our main result so
much. Other important systematic effect is the existence of empty halos
which are neglected in our model. This effect is, however, not important
for the lower limit of $f_{\rm dark}$, because this effect only
increases the fraction of dark halo lenses.   

One possible criticism of our result is that the baryon compression
model we use in this paper is too simple. Many of the previous work,
however, have not addressed this problem. For example, most work of
gravitational lensing statistics which use the mass function of dark
halos assumed that circular velocities of galaxies are the same as
virial velocities of dark halos (this corresponds to $\gamma_v=1$ in our
model). But this assumption seems to be invalid
\citep[e.g.,][]{seljak02a}, and since the number of galaxy lenses scales
as $\propto \gamma_v^4$ the deviation from $\gamma_v=1$ should not be
dismissed. This fact is seen even in our simple model where
the connection between galaxy lenses and dark halo lenses sensitively
depends on $\gamma_v$, and is indeed difficult to
be determined. The importance of baryonic compression, however,
means that we can constrain the model of baryonic compression from
observations of the fraction of dark halo lenses. Although the current
sample of gravitational lensing may be too small for this purpose, the
larger lens sample obtained by e.g., SDSS can strongly constrain the
model of baryonic compression as well as the density profile of dark
halos. 

%%%%%%%%%%%%%%%%%%%%%%%%%%%%%%%%%%%%%%%%%%%%%%%%%%%%%%%%%%%%%%%
%%%%%%%%%%%%%%%%%%%%%%%%%%%%%%%%%%%%%%%%%%%%%%%%%%%%%%%%%%%%%%%
\acknowledgments
%%%%%%%%%%%%%%%%%%%%%%%%%%%%%%%%%%%%%%%%%%%%%%%%%%%%%%%%%%%%%%%
%%%%%%%%%%%%%%%%%%%%%%%%%%%%%%%%%%%%%%%%%%%%%%%%%%%%%%%%%%%%%%%
The author would like to thank Yasushi Suto, Atsushi Taruya, and Mamoru
Shimizu for useful discussions and comments.
%%%%%%%%%%%%%%%%%%%%%%%%%%%%%%%%%%%%%%%%%%%%%%%%%%%%%%%%%%%%%%%%%%%%%%%
%\clearpage
%%%%%%%%%%%%%%%%%%%%%%%%%%%%%%%%%%%%%%%%%%%%%%%%%%%%%%%%%%%%%%%%%%%%%%%

\end{document}